# The doping phase diagram of $Y_{1-x}Ca_xBa_2(Cu_{1-y}Zn_y)_3O_{7-\delta}$ from transport measurements: tracking the pseudogap below $T_c$

S.H. Naqib[1,2*], J.R. Cooper[1], J.L. Tallon[2], R.S. Islam[1], and R.A. Chakalov[3]

[1]IRC in Superconductivity, University of Cambridge, Madingley Road, Cambridge CB3 OHE, UK

[2]MacDiarmid Institute for Advanced Materials and Nanotechnology, Victoria University and Industrial Research Ltd.,
P.O. Box 31310, Lower Hutt, New Zealand

[3]School of Physics and Astronomy, University of Birmingham, Birmingham B15 2TT, UK

## Abstract

The effects of planar hole concentration, p, and magnetic field on the resistivity, $\rho(T)$, of high-quality c-axis oriented crystalline thin films and sintered $Y_{1-x}Ca_xBa_2(Cu_{1-y}Zn_y)_3O_{7-\delta}$ samples were investigated over a wide range of Ca, Zn, and oxygen contents. Zn was used to suppress superconductivity and this enabled us to extract the characteristic pseudogap temperature, $T^*(p)$ below $T_{c0}(p)$ [ $\equiv T_c$ (x =0, y = 0)]. We have also located the characteristic temperature, $T_{scf}$, marking the onset of significant superconducting fluctuations above $T_c$, from the analysis of $\rho(T,H,p)$ and $\rho(T,p)$ data. This enabled us to identify $T^*(p)$ near the optimum doping level where the values of $T^*(p)$ and $T_{scf}(p)$ are very close and hard to distinguish. We again found that $T^*(p)$ depends only on the hole concentration p, and not on the level of disorder associated with Zn or Ca substitutions. We conclude that (i) $T^*(p)$ (and therefore, the pseudogap) persists below $T_{c0}(p)$ on the overdoped side and does not merge with the $T_{c0}(p)$ line and (ii) $T^*(p)$ and the pseudogap energy extrapolate to zero at the doping p = 0.19 ± 0.01.

# 1. INTRODUCTION

One of the most remarkable phenomena in high-$T_c$ cuprates is the observation of the pseudogap in the spectra of charge and spin excitations [1-3]. The pseudogap (PG) is detected by various experimental techniques [1-3] over a certain range of planar hole concentrations, p (the number of added holes per $CuO_2$ plane), extending from the underdoped (UD) to the slightly overdoped (OD) regions. In the 'pseudogap state' various anomalies are observed both in the normal and superconducting (SC) states, which can be interpreted in terms of a depletion of the single particle density of states [3]. A number of theoretical explanations have been proposed as to the origin for the PG, which is believed to be an essential feature of the physics of the normal state (NS) and possibly of the SC state. Existing theories of the PG can be classified broadly into two categories. The first is based upon incoherent Cooper pair formation for $T < T^*$ well above the SC transition temperature, $T_c$, [4-6] with long-range phase coherence appearing only at $T \leq T_c$. In this scenario $T^*$ may be the mean-field $T_c$. $T^*(p)$ merges with the $T_c(p)$ phase curve in the OD region ($p \geq 0.21$) where the pair formation temperature is essentially the same as the phase coherence temperature. In the second scenario, the PG is ascribed to fluctuations of some other type that coexist and generally compete with superconductivity. The most popular picture here is that of short-range antiferromagnetic (AFM) fluctuations, but similar effects have been attributed to charge density waves, a structural phase transition or electronic phase separation on a microscopic scale (e.g., the stripe scenario) [7-10]. One key constraint, which may rule out many of the above models, is that the single-particle density of states has a mysterious states non-conserving property in the PG state [11]. The experimental situation is often thought to be rather inconclusive regarding the origin of the PG [1,2].

The evolution of $\rho(T)$ with p provides a way of establishing the T-p phase diagram of high-$T_c$ superconductors (HTS) and can give estimates of $T^*(p)$, which is identified from a characteristic downturn in $\rho(T)$. One particular difficulty associated this method is that when $p \approx p_{opt}$ (where $T_c(p)$ is maximum), $T^*$ is close to the temperature, $T_{scf}$, at which the effect of SC fluctuation is clearly seen in $\rho(T)$. This is not a problem for theories belonging to the first group where $T^*(p)$ itself is essentially derived from strong

SC fluctuations. For the second scenario, this poses a problem as SC fluctuations (and superconductivity itself) mask the signatures of the predicted PG in the vicinity of (and below) $T_c$. With the notable exception of specific heat measurements [3,11], most experimental techniques are unable to track the predicted $T^*(p)$ below $T_c(p)$. One way of avoiding this is to suppress superconductivity with a magnetic field and reveal the 'NS' below $T_c$, because the PG is comparatively insensitive to magnetic field [12]. In practice this is very hard to accomplish because of the large upper critical field, $H_{c2}$, of the hole-doped HTS. The other way is to destroy superconductivity by introducing disorder, e.g., by alloying with Zn. Because of the d-wave order parameter, Zn substitution suppresses superconductivity most effectively and, like a magnetic field, has little effect on $T^*$ irrespective of hole concentration [13]. We took this second route to look for $T^*(p)$ below $T_{c0}(p)$, though we also employ magnetic fields to help distinguish between SC fluctuations and PG effects.

In this paper we report a systematic study of the transport properties of the superconducting compound $Y_{1-x}Ca_xBa_2(Cu_{1-y}Zn_y)_3O_{7-\delta}$. We have measured resistivity, room-temperature thermopower, S[290K], and the AC susceptibility (ACS) of a series of c-axis oriented thin films and sintered $Y_{1-x}Ca_xBa_2(Cu_{1-y}Zn_y)_3O_{7-\delta}$ samples with different levels of Zn, Ca, and oxygen contents. The motivation for the present study is to locate $T^*(p)$ from the evolution of $\rho(T)$ of Y123 with different amounts of Zn and Ca extending from UD to OD states. While pure Y123 with full oxygenation ($\delta = 0$), is slightly overdoped, further overdoping can be achieved by substituting $Y^{3+}$ by $Ca^{2+}$. The advantages of using Zn are: (i) it mainly substitutes the in-plane Cu(2) sites, thus the effects of planar impurity can be studied and (ii) the doping level remains nearly the same when Cu(2) is substituted by Zn, enabling one to look at the effects of disorder on various normal and SC state properties at almost the same hole concentration [14,15]. From the analysis of $\rho(T,p)$ and $\rho(T,p,H)$ data we extracted the p-dependence of the characteristic temperatures $T^*$ and $T_{scf}$, and find that indeed $T^*$ falls below $T_{c0}$ providing strong evidence for the second scenario, namely that the PG arises from a correlation independent of, and coexisting with, superconductivity.

## 2. EXPERIMENTAL DETAILS AND RESULTS

Single-phase polycrystalline sintered samples of $Y_{1-x}Ca_xBa_2(Cu_{1-y}Zn_y)_3O_{7-\delta}$ were synthesized by solid-state reaction methods using high-purity (> 99.99%) powders. The details of sample preparation and characterization can be found in refs. [16,17]. High-quality c-axis oriented thin films were grown on (*100*) $SrTiO_3$ substrates using the pulsed laser deposition (PLD) method. Details of the PLD parameters and characterization of the films can be found in ref. [17].

The NS and SC properties, including $T^*$, of HTS are highly sensitive to p and, therefore, it is important to determine p as accurately as possible. The room temperature thermopower, S[290K], has a systematic variation with p for various HTS over the entire doping range extending from very UD to heavily OD regimes [18], also S[290K] is insensitive to in-plane disorder like Zn and the crystalline state of the sample [19]. For these reasons S[290K] is a good measure of p even in the presence of strong in-plane electronic scattering by $Zn^{2+}$ ions. For all our samples we have used $S[290K]^{\otimes}$ [20] to determine p. Using these values of p, we find that the parabolic $T_c(p)$ formula [21], given by

$$\frac{T_c(p)}{T_c(p_{opt})} = 1 - Z(p - p_{opt})^2 \tag{1}$$

is obeyed for all samples. Usually, for Zn-free samples, Z and $p_{opt}$ take universal values of 82.6 and 0.16 respectively [21], but these parameters increase systematically with increasing Zn content [22]. In the present case, Z increases from the usual 82.6 for the Zn-free sample to 410 for the 6%-Zn sample [22] and $p_{opt}$ rises from 0.16 for the Zn-free compound to 0.174 for the 6%Zn-20%Ca sample [22]. The physical meaning of these changes in the fitting parameters is that as the Zn content is increased, the region of superconductivity shrinks and becomes centred on higher values of p, before finally

---

$^{\otimes}$S[290K] = -139p + 24.2        for p > 0.155

  S[290K] = 992exp(-38.1p)       for $0.155 \geq p > 0.05$

forming a small "bubble" around p ~ 0.19 and disappearing completely for y ~ 0.1 [17,22,23]. From the observed evolution of $p_{opt}(y)$ we previously inferred that superconductivity for this system is at its strongest at p ~ 0.185 [17,22], as this remains the last point of superconductivity at a critical Zn concentration (defined as the highest possible Zn concentration for which superconductivity just survives, considering all p-values). This point has been made earlier in other studies [23] and the value p ~ 0.19 is indeed a special value at which the PG vanishes quite abruptly, as seen from the analysis of specific heat data of HTS cuprates [11,23].

The hole concentration was varied by changing both the oxygen deficiency and the Ca content. We have obtained $T_c$ from both resistivity and low field ACS measurements ($H_{rms}$ = 0.1Oe; f = 333.3Hz). $T_c$ was taken at zero resistance (within the noise level of $\pm 10^{-6}$ $\Omega$) and at the point where the line drawn on the steepest part of the diamagnetic ACS curve meets the T-independent base line associated with the negligibly small NS signal. $T_c$-values obtained from these two methods agree within 1K for most of the samples [22]. We placed particular emphasis on the determination of p and $T_c$ as accurately as possible because of the extreme sensitivity of various ground-state SC and NS properties to p. This is especially true near the optimum doping level, where, although $T_c(p)$ is nearly flat, the SC condensation energy, superfluid density, PG energy scale and other quantities change quite abruptly and substantially for a small change in p [11,23-25].

Patterned thin films with evaporated gold contact pads and high density (89 to 93% of the theoretical density) sintered bars were used for resistivity measurements. Resistivity was measured using the four-terminal method with an ac current of 1 mA (77 Hz), using 40 $\mu$m dia. copper wire and silver paint to make the low resistance contacts. We have tried to locate the PG temperature, $T^*$, with a high degree of accuracy. At this point we would like to emphasise that $T^*(p)$ does not represent a phase transition temperature but instead $k_B T^*(p)$ is some kind of a characteristic energy scale of a lightly-doped Mott-insulator, probably reflecting the energy of correlated holes and spins [11]. As discussed previously [22] plots of $d\rho(T)/dT$ vs. T and $[\rho(T) - \rho_{LF}]$ vs. T yield very similar $T^*$ values (within $\pm$ 5K) ($\rho_{LF}$ is a linear fit $\rho_{LF}$ = b+cT, in the high-T linear region of $\rho(T)$, above $T^*$). Actually $d\rho(T)/dT$ vs. T is more useful in the sense that it gives a more general measure for the $T^*(p)$ (characterized by an increase in the slope of $\rho(T)$ in the vicinity of $T^*$), than the deviation from the high-T linear $\rho(T)$. Using $T^*/T$ as a scaling parameter, it is also possible to normalize $\rho(T)$. The result of the scaling was shown in ref. [22], where, leaving aside the SC transitions, all resistivity curves collapsed reasonably well on to one p-independent universal curve over a wide temperature range.

It is important to note that Zn does not change the $T^*(p)$ values but it suppresses $T_c(p)$ very effectively. For example, $T^*(p \sim 0.115) \sim 250 \pm 5K$ for both Zn-free and the 3%-Zn samples but $T_c$ itself is suppressed from 70K (Zn-free) to 29K (3%-Zn) [22]. Similar results have been obtained for Zn-substituted Y123 by other studies [13]. This very different Zn-dependence for $T_c$ and $T^*$ has often been stated as an indication for the non-SC origin of the PG [26].

Magneto-transport measurements were made using an *Oxford Instruments* superconducting magnet system. The sample temperature was measured using a *Cernox* thermometer with an absolute accuracy of $\sim 50mK$. For thin film samples magnetic field was always applied along the c-axis, perpendicular to the plane of the film. We have located $T_{scf}$, from the analysis of $\rho(T,H)$ and $d^2\rho(T)/dT^2$ vs. T data, in the T-range from $T_c$ to $\sim T_c + 30K$. Here we have used the facts that (i) $\rho(T,H)$ becomes strongly field sensitive only below $T_{scf}$, where experimentally we find that conventional strong SC fluctuations set in quite abruptly, but $T^*$ is unaffected by magnetic field, at least for H up to 12 Tesla and (ii) $\rho(T)$ shows a much stronger, and progressively accelerating, downturn at $T_{scf}$ than that present at $T^*$. As a consequence, plots of $d^2\rho(T)/dT^2$ vs. T mask the PG feature and visually enhance the effects of SC fluctuations near $T_{scf}$. This is clearly seen in Fig.1. Notice that for nearly identical p ( = 0.134 ± 0.004) both the sintered and the thin film samples with different Ca contents have almost the same values for $T^*$ and $\Delta T_{scf}$, where $\Delta T_{scf} = T_{scf} - T_c$, even though the sintered sample has a much larger resistivity due to the percolative effect and larger contributions from the grain boundaries [27]. We have defined $T_{scf}$ from $d^2\rho(T)/dT^2$ as the temperature at which strong and accelerating downturn in $\rho(T)$ becomes evident near $T_c$. Linear fits of $\rho(T)$ in the temperature range from $\sim T_c + 35K$ to $T_c + 25K$ also locates $T_{scf}$ within ± 2.5K (see the insets of Figs.2). Extraction of $T_{scf}$ from the $\rho(T,H)$ data is shown next in Figs.2 for both sintered and thin film samples. $T_{scf}$ from $\rho(T,H)$ is characterized by the onset of strong magnetic field dependence in the $\rho(T,H)$ data. In summary, three different methods (including the linear fit of $\rho(T)$ to locate the onset of the strong deviation in $\rho(T)$ from linearity near the superconducting transition) are applied and all of them give the same value of $T_{scf}$ to within $\pm 2.5K$.

Fig.3 shows $T_c(p)$ and $T_{scf}(p)$ of two representative sintered 20%-Ca samples with 0%-Zn and 4%-Zn. $\Delta T_{scf}(p)$ is insensitive to Zn content (y) (see the inset of Fig.3), and is also found to be independent of Ca content (x) and crystalline state of the samples (see the inset of Fig.3).

We have tried to model the evolution of $\Delta T_{scf}(p)$ roughly from the temperature dependence of $\rho$ at high temperatures and from the Aslamazov - Larkin (AL) contribution to the fluctuation conductivity [28]. We show this in Fig.4. The excess conductivity due to superconducting fluctuation, $\Delta\sigma(T)$, can be defined as, $\Delta\sigma(T) = 1/\rho(T) - 1/\rho_{BG}(T)$, where, $\rho_{BG}(T)$ is the background resistivity taken as the high-T fit (T > $T_c + 50K$) of the $\rho(T)$ data. The AL contribution to the fluctuation conductivity for a two-dimensional superconductor, $\Delta\sigma^{AL}$, is expressed as, $\Delta\sigma^{AL} = [e^2/(16\hbar d)]\varepsilon^{-1}$ [28], where $\varepsilon = \ln(T/T_c)$ and d is the periodicity of the superconducting layers (d ~ 11.7Å for Y123, assuming that the $CuO_2$ bilayers fluctuate as one unit). In this analysis we have taken $T_c$ at zero resistivity (within the noise level). Fig.4 shows a comparison of the measured T-dependence of $d^2\rho(T)/dT^2$ and that of $d^2\rho_{tot}(T)/dT^2$, where $1/\rho_{tot}(T) = 1/\rho_{BG}(T) + \Delta\sigma^{AL}(T)$. We see that $d^2\rho(T)/dT^2$ and $d^2\rho_{tot}(T)/dT^2$ show remarkable agreement (not surprisingly, even better agreement is seen for thin-film data) despite various uncertainties – for example, theoretical problems regarding an energy or a momentum cut-off [29,30], extrinsic effects due to the grain boundary resistance in polycrystalline samples (which is included in $\rho_{BG}$), and difficulties associated with proper identification of the mean-field SC transition temperature for samples with finite transition width. We have made some model calculations suggesting that as long as the grain boundary resistance remains temperature independent, it does not affect the second derivative significantly.

In an effort to understand the systematic trends in $\Delta T_{scf}$ shown in the inset to Fig. 3 we have examined various terms in the second temperature derivative of $\rho_{tot}(T)$ as follows. Writing

$$\frac{1}{\rho_{tot}(T)} = \frac{1}{\rho_{BG}(T)} + C\varepsilon^{-1} \qquad (2)$$

where $C = e^2/(16\hbar d)$, we obtain (ignoring the insignificant terms)

$$\frac{d^2\rho_{tot}}{dT^2} \approx C\left(\frac{\rho_{tot}}{\varepsilon}\right)^2\left(\frac{1}{T^2}\right)\left\{-\frac{2}{\varepsilon}-1\right\}$$

now, assuming $-\eta$ to be the threshold value of the second derivative for which we can identify a characteristic temperature $T = T_{scf}$ then the above equation becomes, after some rearrangements

$$\frac{\varepsilon_{scf}^3}{2+\varepsilon_{scf}}T_{scf}^2 \approx \frac{C\rho_{scf}^2}{\eta} \qquad (3)$$

where, $\varepsilon_{scf} = \ln(T_{scf}/T_c)$ and $\rho_{scf} = \rho_{tot}(T_{scf})$. We have plotted the left-hand side of equation (3) versus experimental values of $\rho^2(T_{scf})$ in Fig.5 using the approximation $\rho_{tot}(T_{scf}) \sim \rho(T_{scf})$. A remarkably linear trend is found and this gives further credence to our fluctuation analysis. The decreasing trend in $\Delta T_{scf}$ with increasing doping shown in Fig. 3 is therefore primarily associated with the decreasing absolute resistivity as summarised by eq. (3). It is fair to say that $d^2\rho(T)/dT^2$ near $T_c$ is quite insensitive to different scenarios [28-31] for SC fluctuations which may extend the high-$T$ fluctuation range up to $3T_c$ or more. Our aim was only to study the strong SC fluctuation near the transition temperature so that $T^*$ can be distinguished from $T_{scf}$. The presence of decreasing fluctuation conductivity at higher temperatures does not affect our analysis in any significant way. A more rigorous recent analysis of thin film data has also found the paraconductivity to be independent of the PG [31].

Once we have located $T_{scf}$, we are in a position to look at $T^*(p)$ (below $T_{c0}(p)$) for Zn substituted samples. There is one disadvantage of Zn substitution that can hamper the identification of $T^*$ from $\rho(T)$ measurements, namely, Zn tends to localize low-energy quasiparticles (QP) and induces an upturn in $\rho(T)$. This upturn starts at increasingly higher temperatures as p decreases and, to a lesser extent, as the Zn content increases, and thus can mask the downturn due to the PG at $T^*$ [13,22]. Indeed, the upturn temperature, $T_{min}$, has been found to scale with $T^*$ and is evidently also associated with the pseudogap, trending to zero as $p \rightarrow 0.19$ [32]. In this study we have taken care of this fact by confining our attention to the $\rho(T,p)$ data for lower Zn contents in the underdoped region (p < 0.14). Fig.6a shows the $\rho(T)$ data of sintered $Y_{0.80}Ca_{0.20}Ba_2(Cu_{0.96}Zn_{0.04})_3O_{7-\delta}$ for p = $0.174 \pm 0.003$. The insets of Fig.6 clearly show the downturn associated with the PG at around 80K. It should be noted that $T_c$ of this compound is 43K and $T_{scf}$ = 62.5K (see Fig.3). Fig.6b shows a similar analysis for sintered $Y_{0.80}Ca_{0.20}Ba_2(Cu_{0.985}Zn_{0.015})_3O_{7-\delta}$ with p = $0.177 \pm 0.003$. The $\rho(T)$ features of this compound once again show a $T^* \sim 80K$ with now a $T_c$ of 54K and $T_{scf}$ = 73K. Considering the facts that $T^*(p)$ values are the same irrespective of Ca and Zn contents (at least for the level of substitutions used in the present study)[17,22], and $T_{c0-max}$ for pure Y123 is 93K, these $T^*(p)$ are substantially below the respective $T_{c0}(p)$ values ($\sim$ 90.5K) (see Fig.8). A similar result is shown for $Y_{0.95}Ca_{0.05}Ba_2(Cu_{0.98}Zn_{0.02})_3O_{7-\delta}$ thin film with p = 0.167 $\pm$ 0.003, $T_c$ = 63.6K, and $T_{scf}$ = 80K in Fig.6c. Notice that $T^*$ = 92.5K of this film almost coincides with its $T_{c0}(p) \sim 92K$. In the insets, the blue shaded parts show the range of regions of extremely strong SC fluctuations extending down from $T_{scf}$ to $T_c$. For comparison the error bars on $T^*$ are shown by the yellow shaded regions. This illustrates that these two regions can be clearly distinguished, and that they both fall significantly below $T_{c0}$. This study, to our knowledge, is the first instance where $T^*(p)$ has been tracked down below $T_{c0}(p)$ from any transport measurement, although of course this has effectively been done earlier by analysis of specific heat data by Loram *et al.* [3,11] and NMR data by Tallon *et al.* [33].

The field dependence of $T^*$ and $T_{scf}$ for two thin film samples with different values of p (one is underdoped and the other is slightly overdoped) are shown in Figs.7. It is clearly seen that $T^*(p)$ is completely field independent up to 12 Tesla while $T_{scf}(p)$ is strongly field dependent and shifts to lower values with increasing magnetic field just as $T_c$ itself does. A magnetic field of 6 Tesla decreases $T_{scf}$ by (5 $\pm$ 1)K for the two compounds shown in Fig.7. This clearly relates $T_{scf}$ to superconductivity and $T^*$ to a different type of correlation. In Fig.8 we construct a doping "phase diagram" for $Y_{1-x}Ca_xBa_2(Cu_{1-y}Zn_y)_3O_{7-\delta}$, including the PG energy scale, $E_g(p)/k_B$, from specific heat measurements [23]. $E_g$ is the energy scale for the PG and $E_g(p)/k_B \sim \theta T^*(p)$, where ($\theta = 1.3 \pm 0.1$) is a certain proportionality constant [26]. This figure clearly shows that $T^*$ and $T_{c0}$ have very different doping dependence, most importantly so in the overdoped regions. In particular $T^*(p)$ does not merge with $T_c$ in the overdoped region but goes to zero at $p = 0.19 \pm 0.01$, following the same trend as found for $E_g(p)/k_B$.

# 3. DISCUSSION

From a careful analysis of the resistivity data we have been able to track $T^*(p)$ below $T_{c0}(p)$. At this point we would like to stress once again that $T^*(p)$ values obtained from $\rho(T)$ are independent of the crystalline state and are the same for polycrystalline and single crystal samples [22]. Considering the disorder (Zn and Ca) independence of $T^*(p)$, our findings confirm that the PG exists in the SC region below $T_{c0}$ and $T^*(p)$ does not merge with $T_c(p)$ in the OD region as proposed by various precursor pairing scenarios for PG. As explained in the present work, if one is unable to distinguish between $T^*(p)$ and $T_{scf}(p)$, one can easily be led to the wrong conclusion that $T^*(p)$ exists for $p > 0.19$ and merges with $T_c(p)$ on further overdoping. As $T_{scf}$ is $\sim T_c + 22K$ near optimum doping level, the close proximity between $T_{scf}$ and $T^*$ makes an accurate identification of $T^*$ very difficult unless both $T_c$ and thus $T_{scf}$ are suppressed by some means that does not affect $T^*$. Zn serves this purpose very well. Unfortunately, we have been unable to track $T^*(p)$ for samples with $p > 0.180$ with enough accuracy [17]. This is because $T^*(p)$ decreases very sharply with increasing $p$ (compared with the decreases in $T_c(p)$ and $T_{scf}(p)$, see Figs.3 and 8) and becomes very close to or even goes below $T_{scf}$ and eventually $T_c(p)$ [3,11]. We discuss further the following examples (see Figs.9): (i) $Y_{0.90}Ca_{0.10}Ba_2(Cu_{0.985}Zn_{0.015})_3O_{7-\delta}$ with $p = 0.180 \pm 0.003$ has a $T_c = 64K$ and $\rho(T)$ is linear over 310K to 80K, then a marked and accelerating downturn in $\rho(T)$ starts at $\sim$ 79K. Considering the Zn and Ca independence of $\Delta T_{scf}(p)$ and $T^*(p)$, we expect $T_{scf}$ to be $T_c + \Delta T_{scf} \approx (64 + 17)K = 81K$, where $\Delta T_{scf} \approx 17K$ is read off the inset of Fig.3. So this downturn at ~ 79K must be the onset of strong SC fluctuations. Consequently $T^*$ must lie below 79K in this sample (ii) Another example is $Y_{0.80}Ca_{0.20}Ba_2(Cu_{0.96}Zn_{0.04})_3O_{7-\delta}$ with p = 0.184 ± 0.003, which has $T_c$ = 42.25K and ρ(T) is linear over 310K to 60K. A marked and accelerating downturn in ρ(T) starts at ~ 58.5K. We expect $T_{scf}$ to be $T_c + \Delta T_{scf} \approx$ (42.25 + 16.5)K = 58.75K for this sample. Thus for this compound $T^*$ < 58.5K, much less than $T_{c0}$ ~ 89K.

$T_{scf}$(p) has been taken as the onset temperature of strong SC fluctuations in the present study. The disorder independence of $\Delta T_{scf}$(p) (inset of Fig.3), a simple AL-type analysis of ρ(T) data (see Figs.4 and 5), and the magnetic field dependence of $T_{scf}$ strongly support this assumption. Disorder and magnetic field suppress both $T_c$(p) and $T_{scf}$(p) in the same qualitative way, unlike $T^*$(p) which remains unaffected.

## 4. CONCLUSIONS

In summary, we have analysed resistivity data to determine the $T^*$(p) of $Y_{1-x}Ca_xBa_2(Cu_{1-y}Zn_y)_3O_{7-\delta}$ over a wide doping range and compositions. We have shown that $T^*$(p) falls below $T_{c0}$(p) in the overdoped side, does not merge with $T_c$(p), and the extrapolated $T^*$(p) becomes zero at p = 0.19 ± 0.01. We have also extracted a characteristic temperature, $T_{scf}$, at which strong SC fluctuations become detectable. It is perhaps surprising that $T_{scf}$ is so well-defined experimentally in all our samples, but the very different p, H, and Zn dependence of $T^*$ and $T_{scf}$ points towards their different origins, e.g., $T_{scf}$ is related to precursor SC, whereas $T^*$ has a non-SC origin. Our findings support the picture proposed by Loram *et al.* based on their specific heat study that the PG vanishes at a critical doping, $p_{crit}$ ~ 0.19, and coexists with SC for p < 0.19 [11,23]. Recently Krasnov *et al.* have reached similar conclusions based on their intrinsic tunneling spectroscopy studies of Bi-2212 single crystals [12,34].

## ACKNOWLEDGEMENTS

We gratefully acknowledge Dr. J.W. Loram for helpful comments and suggestions. We also thank Dr. J. Durrell for his help with patterning the films. SHN acknowledges the financial support from the Commonwealth Scholarship Commission (UK), Darwin

College, Cambridge Philosophical Society, Lundgren Fund, and the Department of Physics, Cambridge University. RSI acknowledges the financial support from Trinity College, University of Cambridge, and the Cambridge Commonwealth Trust (UK). Thanks are also due to the New Zealand Marsden Fund (JLT).

\* Corresponding author. Email address: s.naqib@irl.cri.nz

**Figure Captions (S.H. Naqib *et al.*)**

**[PAPER TITLE: The doping phase diagram of ..........]**

Fig.1. Resistivity data for (a) $Y_{0.95}Ca_{0.05}Ba_2Cu_3O_{7-\delta}$ ($T_c$ = 81.8K, p = 0.133, $T_{scf}$ = 105K, and $T^*$ = 210K) thin film and (b) $Y_{0.80}Ca_{0.20}Ba_2Cu_3O_{7-\delta}$ ($T_c$ = 81K, p = 0.136, $T_{scf}$ = 107K, and $T^*$ = 202K) sintered sample. Insets show the identification of $T^*$ and $T_{scf}$. Notice that $d^2\rho(T)/dT^2$ is completely featureless at $T^*$. ● denotes [$\rho(T)$ - $\rho_{LF}$] data and **O** denotes $d^2\rho(T)/dT^2$ data. The dashed-dotted straight lines in the insets are drawn as guides to the eye.

Fig.2. Determination of $T_{scf}$ from $\rho(T,H)$ data. Main panel: [$\rho(H = 6$ Tesla)/ $\rho(H = 0$ Tesla)] vs. T data. Insets: $\rho(T)$ and its linear fit (the dashed-dotted thick straight line) near $T_c$ and $d^2\rho(T)/dT^2$ (a) $Y_{0.80}Ca_{0.20}Ba_2(Cu_{0.97}Zn_{0.03})_3O_{7-\delta}$ (sintered, p = 0.185, and $T_c$ = 45.8K), (b) $Y_{0.80}Ca_{0.20}Ba_2(Cu_{0.96}Zn_{0.04})_3O_{7-\delta}$ (sintered, p = 0.198, and $T_c$ = 36.7K), and (c) $Y_{0.95}Ca_{0.05}Ba_2Cu_3O_{7-\delta}$ (thin film, p = 0.148, and $T_c$ = 85.2K). The straight line in the main panel of Fig.2c is drawn as a guide to the eye.

Fig.3. Main panel: $T_c(p)$ and $T_{scf}(p)$ for sintered $Y_{0.80}Ca_{0.20}Ba_2(Cu_{1-y}Zn_y)_3O_{7-\delta}$. Inset: $\Delta T_{scf}(p)$ for sintered and thin films of $Y_{1-x}Ca_xBa_2(Cu_{1-y}Zn_y)_3O_{7-\delta}$. Zn and Ca contents (x and y) are shown in the figure.

Fig.4. $d^2\rho(T)/dT^2$ and $d^2\rho_{tot}/dT^2$ vs. T for sintered $Y_{0.90}Ca_{0.10}Ba_2Cu_3O_{7-\delta}$. $\rho(T)$ is the experimental resistivity and $1/\rho_{tot}$ is the sum of the background conductivity and the AL fluctuation conductivity (see text). (a) An overdoped sample ($T_c$ = 78K, p = 0.183) and (b) an underdoped sample ($T_c$ = 83K, p = 0.149). Arrows indicate $T_{scf}$. Insets show $\rho(T)$ and its linear fit (dashed-dotted straight line) near the superconducting transition.

Fig.5. [$(\varepsilon_{scf}^3 T_{scf}^2)/(2 + \varepsilon_{scf})$] vs. $\rho^2(T_{scf})$ (see eq. (3) in text) for the sintered $Y_{0.80}Ca_{0.20}Ba_2(Cu_{1-y}Zn_y)_3O_{7-\delta}$ samples. Zn-contents (y in %) are shown in the figure. The dashed straight line is drawn as a guide to the eye.

Fig.6. $T^*$ for (a) sintered $Y_{0.80}Ca_{0.20}Ba_2(Cu_{0.96}Zn_{0.04})_3O_{7-\delta}$ (p = 0.174 ± 0.003), (b) sintered $Y_{0.80}Ca_{0.20}Ba_2(Cu_{0.985}Zn_{0.015})_3O_{7-\delta}$ (p = 0.177 ± 0.003), and (c) thin film of $Y_{0.95}Ca_{0.05}Ba_2(Cu_{0.98}Zn_{0.02})_3O_{7-\delta}$ (p = 0.167 ± 0.003). Top inset shows $d^2\rho(T)/dT^2$ data. Notice that $d^2\rho(T)/dT^2$ is featureless at $T^*$ (yellow region in the bottom inset). $T_{scf}$ is also marked in the bottom inset. The straight lines in the bottom inset are drawn as a guide to the eye. Strong superconducting fluctuations persist in the blue region.

Fig.7. Magnetic field dependence of $T^*$ and $T_{scf}$. (a) A second slightly overdoped $Y_{0.95}Ca_{0.05}Ba_2(Cu_{0.98}Zn_{0.02})_3O_{7-\delta}$ thin film (p = 0.168 ± 0.003 and $T_c$ = 64K) and (b) an underdoped $Y_{0.95}Ca_{0.05}Ba_2Cu_3O_{7-\delta}$ thin film (p = 0.148 ± 0.003 and $T_c$ = 85.2K).

Fig.8. Characteristic pseudogap energies ($T^*$ and $E_g/k_B$) for $Y_{1-x}Ca_xBa_2(Cu_{1-y}Zn_y)_3O_{7-\delta}$: The thick black dashed line shows $T_{c0}(p)$ for pure Y123, drawn using equation (1) with $T_{c0-max}$ = 93K. The thick black straight line is a guide to the eye. The dashed blue line represents the large asymmetric error bar associated with the data point for $Y_{0.80}Ca_{0.20}Ba_2(Cu_{0.96}Zn_{0.04})_3O_{7-\delta}$ (p = 0.184 ± 0.003) compound (see text for details).

Fig.9. Two examples where $T_{scf} \geq T^*$ (see text for details): (a) $Y_{0.90}Ca_{0.10}Ba_2(Cu_{0.985}Zn_{0.015})_3O_{7-\delta}$ (p = 0.180 ± 0.003 and $T_c$ = 64K) and (b) $Y_{0.80}Ca_{0.20}Ba_2(Cu_{0.96}Zn_{0.04})_3O_{7-\delta}$ (p = 0.184 ± 0.003 and $T_c$ = 42.25K).

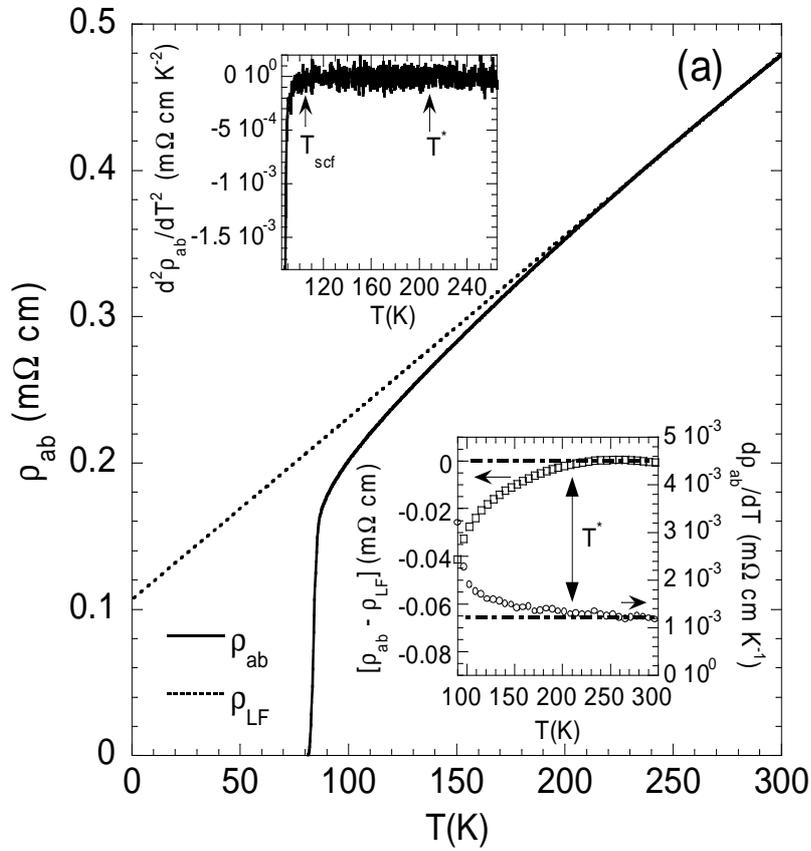

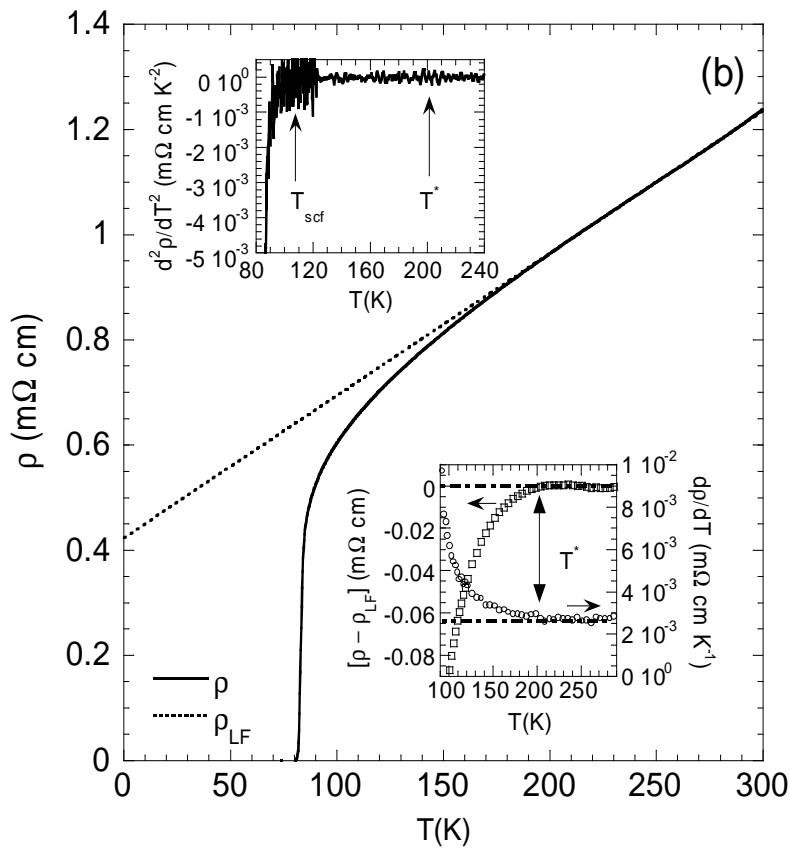

Fig.2 (S.H. Naqib *et al.*) [The doping phase..............]

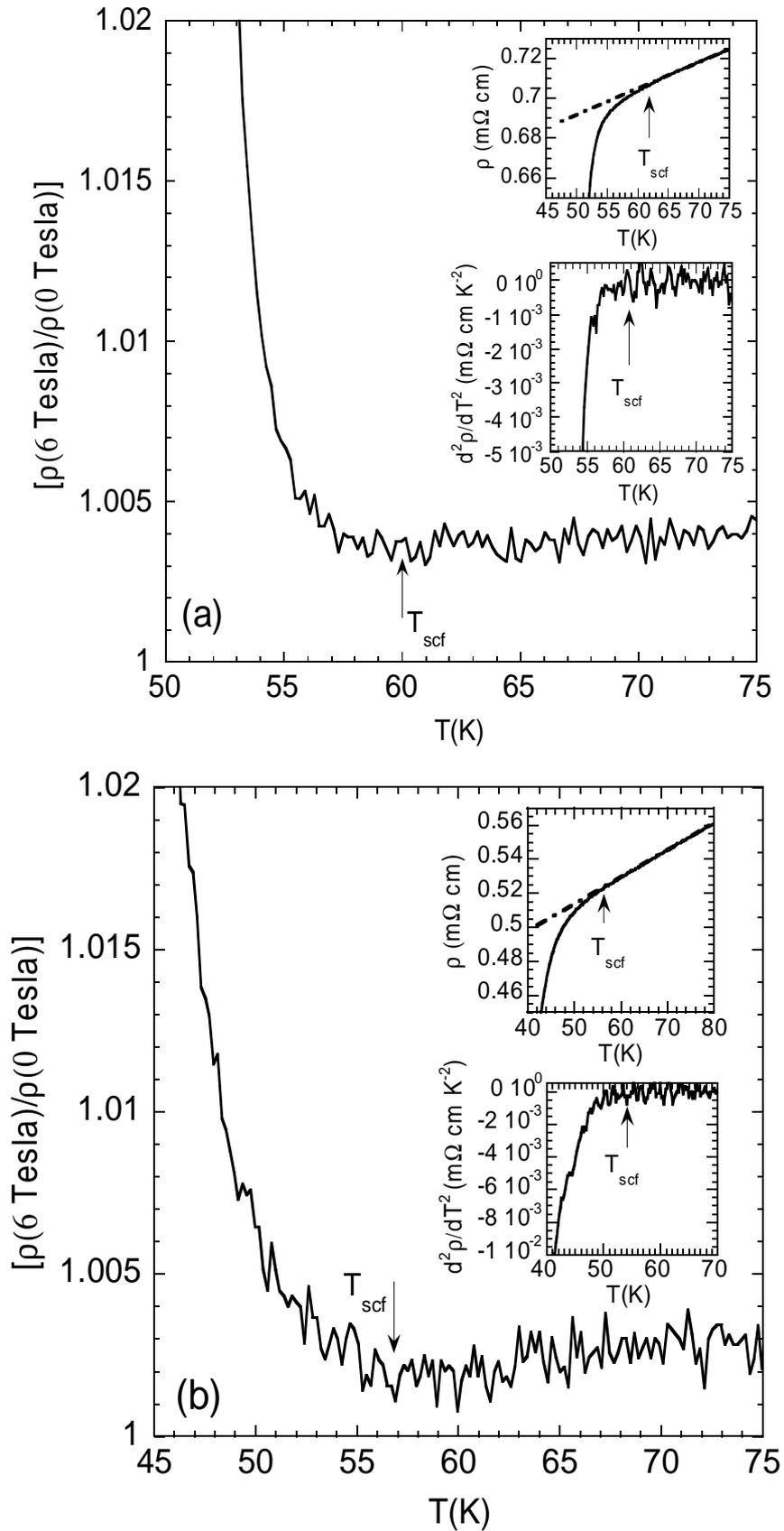

Fig.2 (S.H. Naqib *et al.*) [The doping phase…………..]

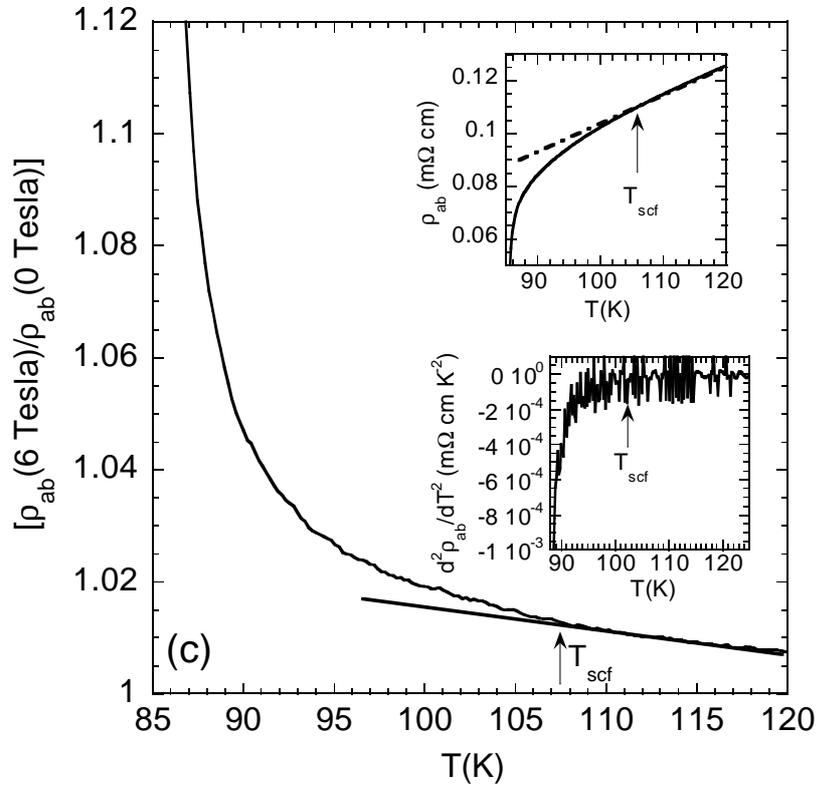

Fig.3 (S.H. Naqib *et al.*) [ The doping phase…………..]

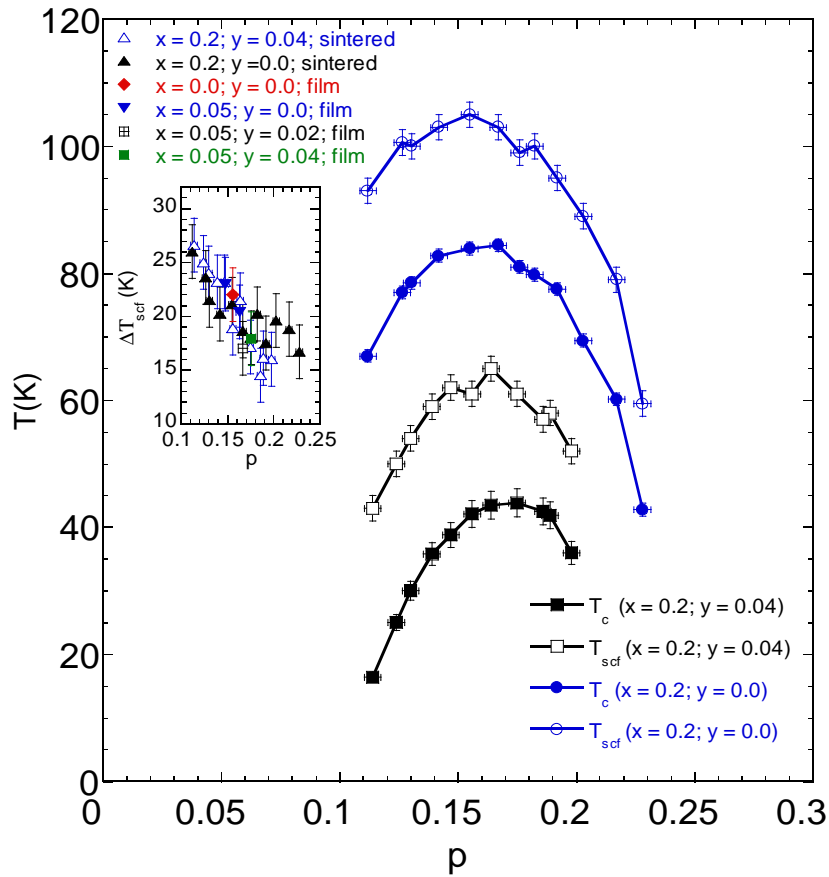

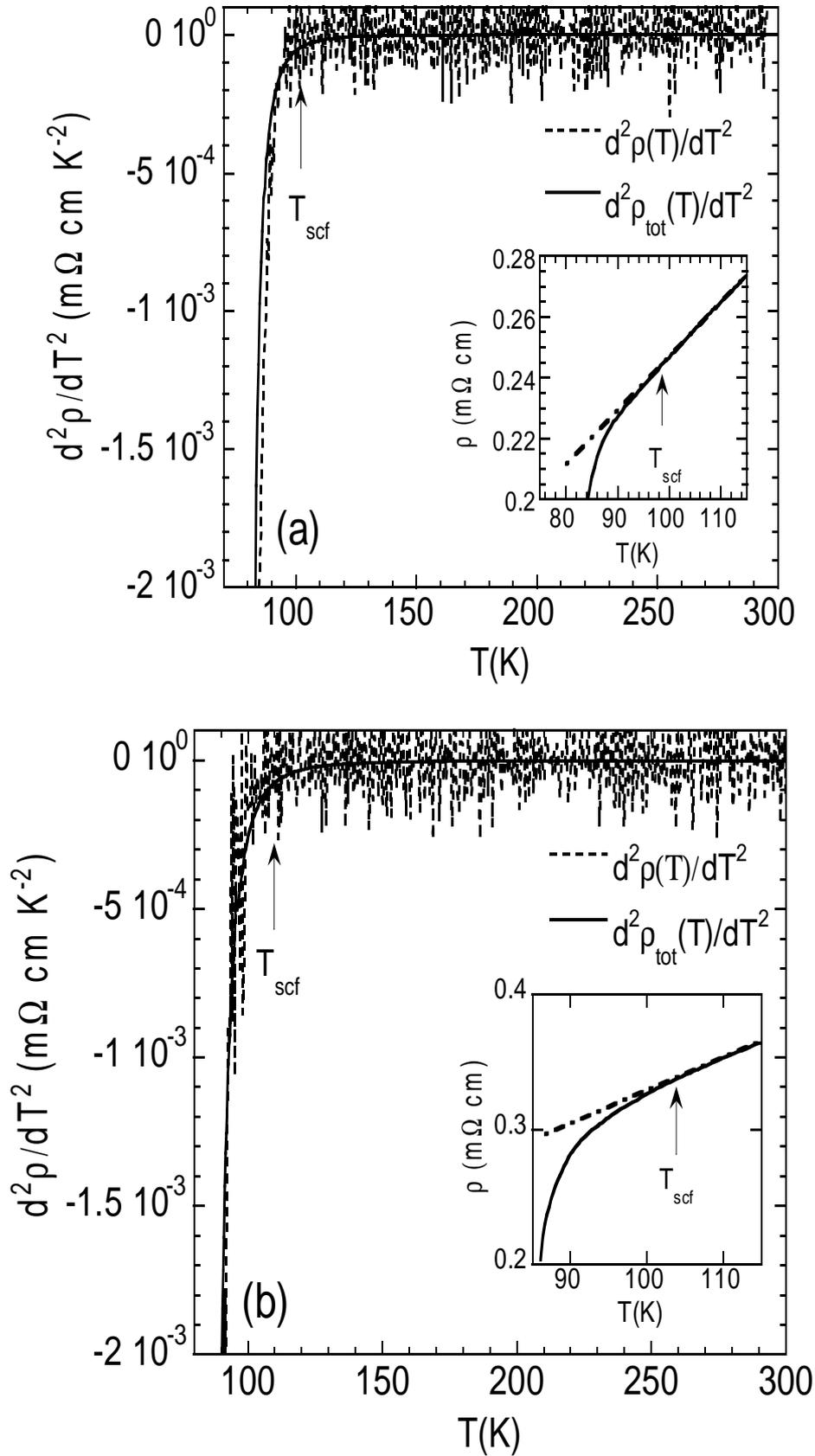

Fig.5 (S.H. Naqib *et al.*) [The doping phase…………..]

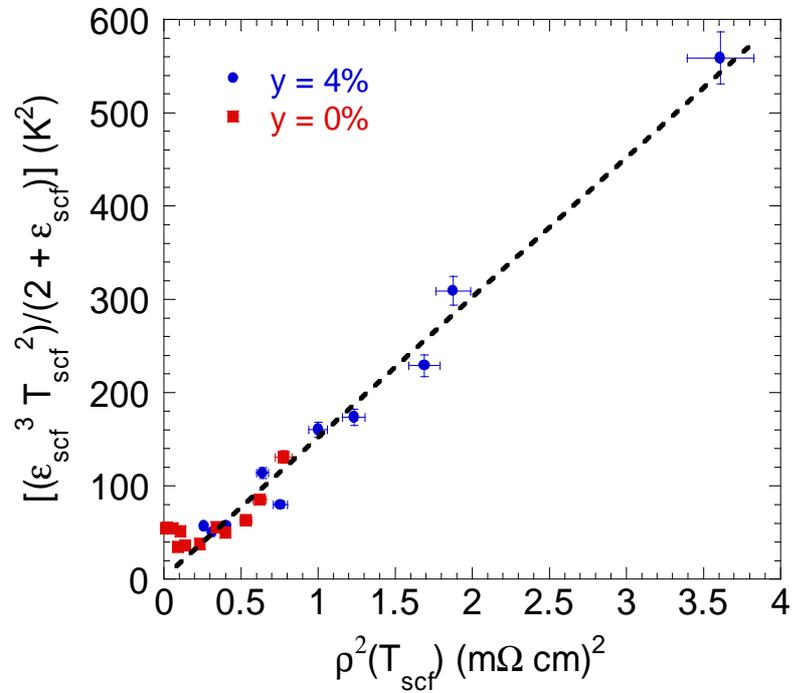

Fig.6 (S.H. Naqib *et al.*) [The doping phase…………..]

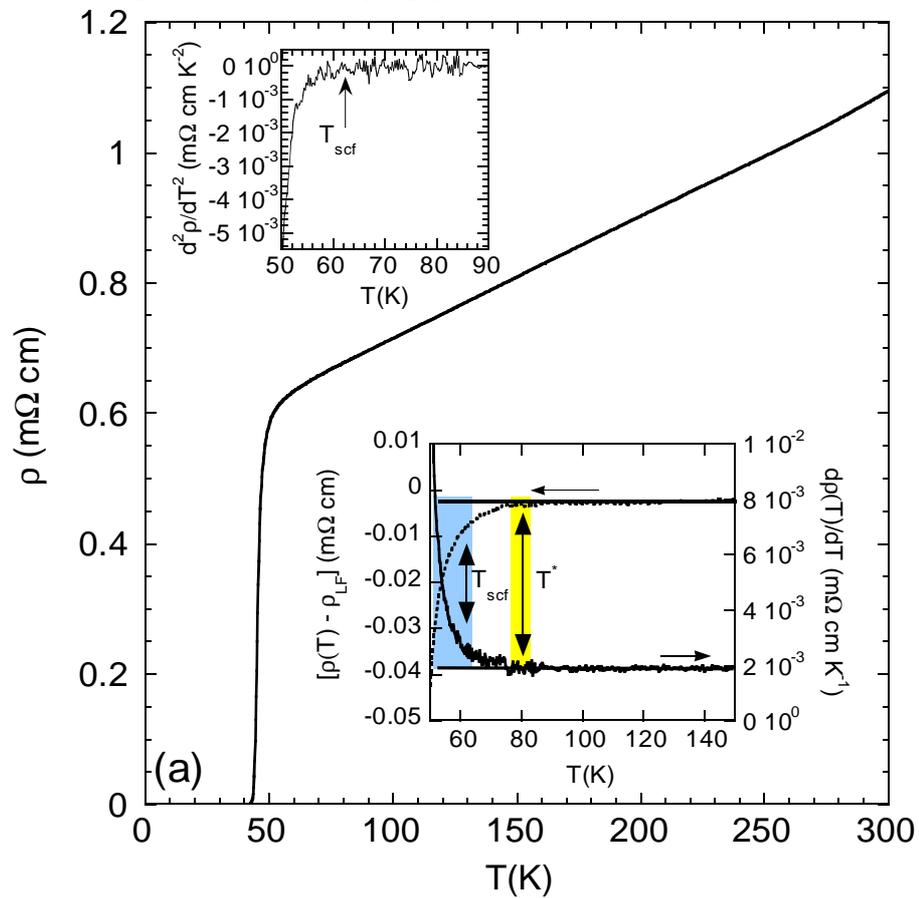

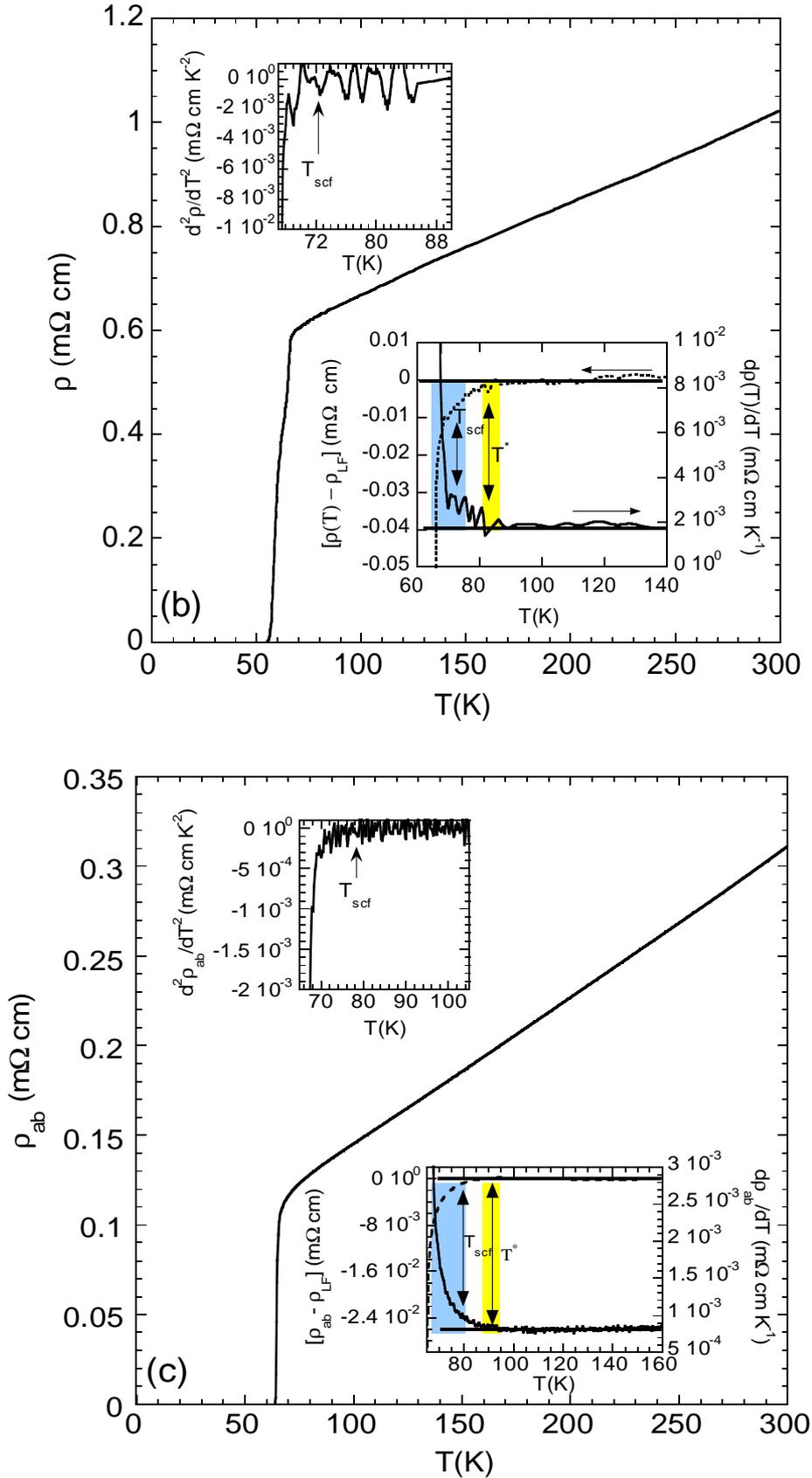

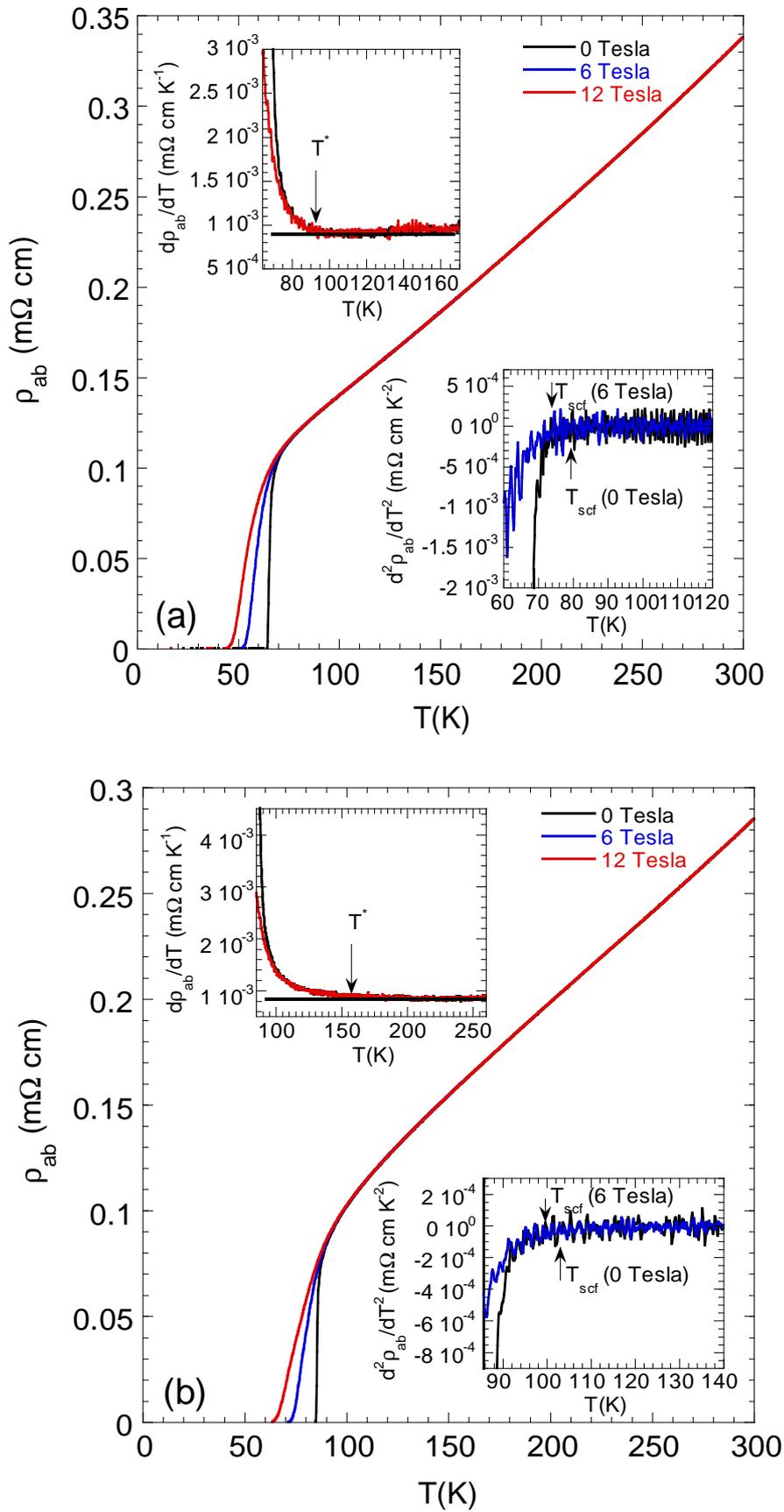

Fig.8 (S.H. Naqib *et al.*) [The doping phase.............]

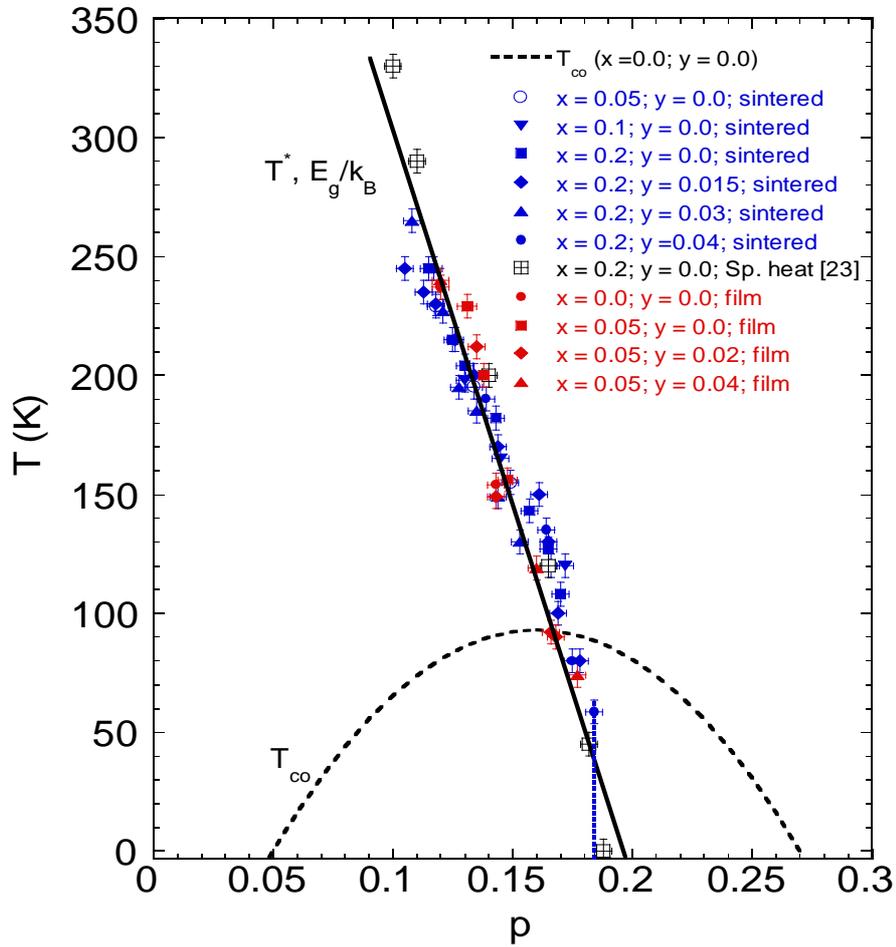

Fig.9 (S.H. Naqib *et al.*) [The doping phase.............]

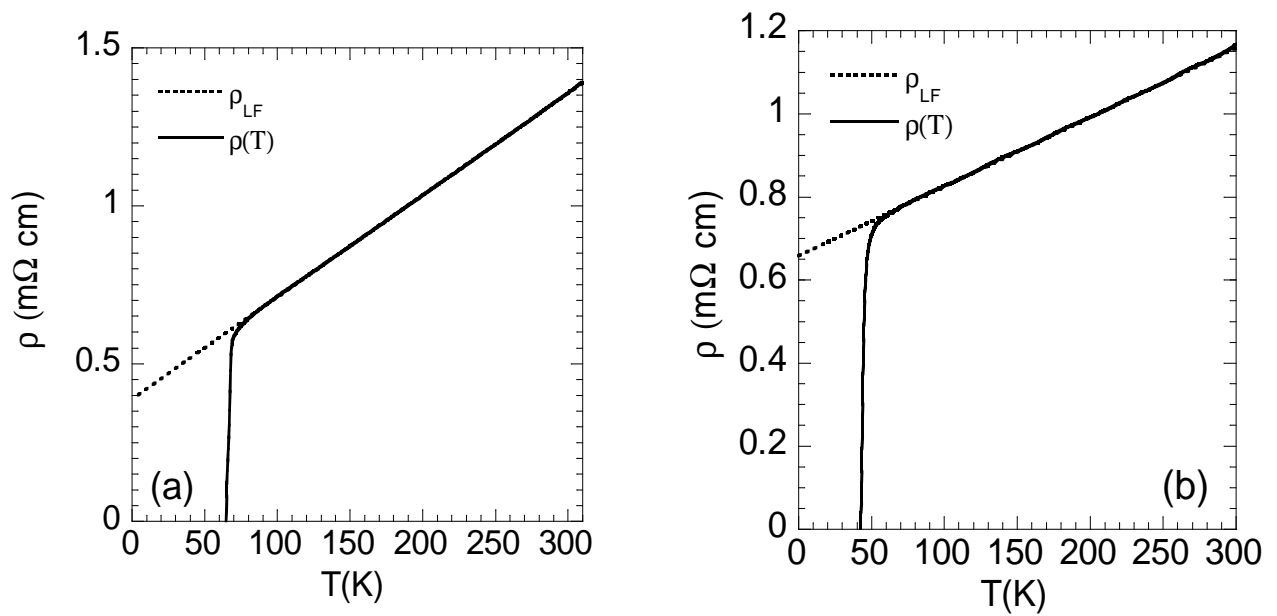